\documentclass[12pt]{article}
\usepackage[utf8]{inputenc}
\usepackage[english]{babel}
\usepackage{amsmath}
\usepackage{mathrsfs}
\usepackage{amssymb}
\usepackage{amsthm}
\usepackage{amsfonts}
\usepackage{xspace}
\usepackage[normalem]{ulem}
\usepackage{graphicx}
\usepackage{multirow}
\usepackage{appendix}
\usepackage{enumerate}
\usepackage{url} 
\usepackage{authblk}
\usepackage{float}
\usepackage{xcolor}
\usepackage{subcaption}

\usepackage{xcolor}
\usepackage{soul}
\usepackage{tikz}
\usetikzlibrary{calc}

\usepackage[top=1in,bottom=1.5in,left=1in,right=1in]{geometry}

\usepackage{natbib}
\usepackage[unicode=true,bookmarks=true,bookmarksnumbered=false,bookmarksopen=false,breaklinks=false,pdfborder={0 0 1},backref=false,colorlinks=true]{hyperref}
\hypersetup{citecolor=blue}
\usepackage{url}

\newcommand{\bX}{\boldsymbol{X}}

\newcommand{\bone}{\boldsymbol{1}}
\newcommand{\bbeta}{\boldsymbol{\beta}}

\title{Randomization Inference for Cluster-Randomized Test-Negative Designs with Application to Dengue Studies\\
\vspace{5pt}
\large Unbiased estimation, Partial compliance, and Stepped-wedge design}

\author[1]{Bingkai Wang}
\author[2]{Suzanne M. Dufault}
\author[1]{Dylan S. Small}
\author[3]{Nicholas P. Jewell}

\affil[1]{\small Department of Statistics and Data Science, the Wharton School, University of Pennsylvania, PA, U.S.A.}
\affil[2]{\small Division of Pulmonary and Critical Care Medicine, University of California, San Francisco, CA, U.S.A.}
\affil[3]{\small Department of Medical Statistics,
London School of Hygiene \& Tropical
Medicine, U.K.}

\begin{document}
\def\spacingset#1{\renewcommand{\baselinestretch}%
{#1}\small\normalsize} \spacingset{1}

\date{\vspace{-5ex}}

\maketitle
\begin{abstract}
In 2019, the World Health Organization identified dengue as one of the top ten global health threats. For the control of dengue, the Applying {\em Wolbachia} to Eliminate Dengue (AWED) study group conducted a cluster-randomized trial in Yogyakarta, Indonesia, and used a novel design, called the cluster-randomized test-negative design (CR-TND). This design can yield valid statistical inference with data collected by a passive surveillance system and thus has the advantage of cost-efficiency compared to traditional cluster-randomized trials. 
We investigate the statistical assumptions and properties of CR-TND under a randomization inference framework, which is known to be robust and efficient for small-sample problems. We find that, when the differential healthcare-seeking behavior comparing intervention and control varies across clusters (in contrast to the setting of \citealp{dufault2020analysis} where the differential healthcare-seeking behavior is constant across clusters), current analysis methods for CR-TND can be biased and have inflated type I error. We propose the log-contrast estimator that can eliminate such bias and improve precision by adjusting for covariates. Furthermore, we extend our methods to handle partial intervention  compliance and a stepped-wedge design, both of which appear frequently in cluster-randomized trials. Finally, we demonstrate our results by simulation studies and re-analysis of the AWED study.

\end{abstract}
\noindent%
\vfill

\newpage
\spacingset{1.5}
\setcounter{page}{1}

\section{Introduction}\label{sec:intro}
\subsection{Motivating example: An intervention to reduce dengue incidence}
Dengue is a widespread, rapidly increasing arboviral disease, primarily transmitted by \textit{Aedes aegypti} mosquitoes.
Every year, there are an estimated 50 million to 100 million dengue cases globally \citep{cattarino2020mapping}.
To reduce dengue transmission, recent virological research has shown that  \textit{Aedes aegypti} mosquitoes that are transinfected with the bacterium, \textit{Wolbachia}, are more resistant to spread arboviral diseases \citep{rainey2014understanding, johnson2015impact, dutra2016wolbachia}. In addition, \textit{Wolbachia}-infected mosquitoes can stably invade wild \textit{Aedes aegypti} mosquito populations through advantageous reproductive outcomes between \textit{Wolbachia}-infected and wild mosquitoes \citep{walker2011w}. Based on these advances, the World Mosquito Program has launched  worldwide studies that have successfully used \textit{Wolbachia}-infected mosquitoes to control for arboviral diseases.

The Applying {\em Wolbachia} to Eliminate Dengue (AWED) study is an unblinded cluster-randomized trial that evaluated the efficacy of {\em Wolbachia}-infected mosquito deployments to reduce dengue incidence in Yogyakarta, Indonesia \citep{utarini2021efficacy}. In this study, twenty-four contiguous geographical clusters were equally randomized to receive intervention (initial {\em Wolbachia}-infected mosquito deployments) or control (no intervention). Intervention is thus assigned at the cluster level, i.e., all individuals in the same cluster are exposed to the same intervention, due to its nature. During a two to three-year follow-up, incident dengue cases were recruited by a study-initiated passive surveillance system: among recruited patients with acute fever who presented at several local primary clinics, a laboratory test for dengue was performed; dengue cases comprised of positive test results, with cases of other febrile illnesses (OFI) providing a natural control group for comparison. This design thus represents an innovative form of the original test-negative design (TND), a modified case-control design traditionally used for vaccine evaluation (\citealp{jackson2013test}, explained in Section~\ref{subsec:CR-TND} below). We refer to this design used by the AWED study as a cluster-randomized test-negative design (CR-TND), following the terminology in \cite{anders2018cluster}.

Distinct from classical cluster-randomized trials, the passive patient recruitment procedure in the AWED study is impacted by  healthcare-seeking behavior of potential patients. That is,  observed case counts only reflect the population of active healthcare seekers instead of the entire population of cluster residents. This is central to the concept of TNDs that are intended to reduce confounding associated with healthcare-seeking behavior \citep{sullivan2016theoretical}.
However, since intervention assignment was not blinded in AWED, healthcare-seeking behavior may be differentially affected by knowledge of intervention assignment. For example, dengue patients in treated clusters might be less concerned with fever symptoms as compared to control cluster inhabitants, leading to fewer clinical visits and subsequently reduced dengue counts in the treated clusters. 
In this article, we explain how CR-TND addresses this issue, describe the statistical assumptions and properties of CR-TND, discuss and extend currently used methods for CR-TND to handle (i) cluster variation in healthcare-seeking behaviors, (ii) partial compliance, and (iii) a stepped-wedge design.

\subsection{Cluster-randomized studies with a TND}\label{subsec:CR-TND}
In cluster-randomized trials where data are passively collected, a CR-TND borrows methods from a standard TND to yield valid inferences.
A TND is a modification of a case-control study popularly used for studying vaccine effectiveness in observational studies, where  observed data are  counts of test-positives (cases) and test-negatives (controls) among  vaccinated and unvaccinated healthcare seekers. It most closely resembles a case-cohort design in concept except that data are passively collected instead of randomly sampled.
Under the assumption that vaccination has no effect on the test-negative conditions, the relative risk among healthcare seekers can be estimated by the odds ratio based on these observed counts; estimation of the  relative risk for the entire population, however, requires an external validity assumption \citep{jackson2013test, haber2015probability}. \cite{sullivan2016theoretical} considered the TND from a causal perspective, and showed that a TND can reduce, but not necessarily eliminate  bias from differential healthcare-seeking behavior in observational studies.

A CR-TND is similar to a TND in assumptions and data structure. The similarity between CR-TND and TND provides a CR-TND with the statistical foundation for estimating the relative risk and, as a result, makes a CR-TND a cost-effective and simple design for studying rare outcomes compared to a prospective design \citep{anders2018cluster}. The major difference between a CR-TND and a TND is the exposure: in a CR-TND, intervention is randomly assigned  at a community (cluster) level, while the exposure in a TND can vary across individuals and be potentially confounded by covariates. Such a difference suggests that, in a CR-TND, the healthcare-seeking behavior exposure is no longer confounded (due to random assignment). However, as for all TNDs, intervention is unblinded in the AWED study, and patients may potentially change their healthcare-seeking behavior given  knowledge of their intervention assignment.

Since \cite{anders2018cluster} first introduced the CR-TND, \cite{jewell2019analysis} proposed a series of cluster-level estimators for the relative risk, basing inference on permutation distributions and approximations thereof. \cite{dufault2020analysis} later addressed the impact of differential healthcare-seeking behavior caused by unblinded intervention assignment by considering a case-only analysis strategy. 
A further challenge of  differential healthcare-seeking behavior occurs when the latter varies across clusters, which is not considered in \cite{dufault2020analysis}. 
Furthermore, it is valuable to account for non-compliance or partial compliance with the assigned intervention. All these issues are illustrated in the AWED study and can potentially occur in future studies involving a CR-TND.

In this article, we address the above open questions. We extend the randomization inference framework for CR-TNDs, thereby avoiding distributional assumptions on the outcomes, with validity even for a small number of clusters. By statistically formalizing the assumptions for CR-TND, we identify and quantify the bias of existing methods (including the odds ratio estimator, test-positive fraction estimator, and model-based estimators) when differential healthcare-seeking behavior between intervention and control arms varies across clusters.
We propose an unbiased estimator for the intervention/treatment relative risk, called the log-contrast estimator, and further incorporate covariate adjustment to improve its precision. In addition, partial compliance is accommodated through an instrumental variable method. 

Recognizing the increasing use of a stepped-wedge design for intervention assignment in cluster-randomized trials \citep{hussey2007design, li2021mixed}, we also extend CR-TNDs to accommodate staggered intervention assignment procedures. For a stepped-wedge CR-TND, we characterize the underlying structure, assumptions, and unbiased estimation methods following a similar strategy for discussing parallel-arm CR-TNDs. Our results build on key results from \cite{ji2017randomization, roth2021efficient}, which propose randomization inference methods for analyzing a traditional stepped-wedge design. We extend their randomization inference framework and, to the best of our knowledge, provide the first statistical characterization and analysis approach for stepped-wedge CR-TNDs.

The remainder of this article is organized as follows. In the next section, we describe our randomization inference framework and present the assumptions for estimation based on a CR-TND. In Section~\ref{sec:existing-estimators}, we briefly review existing estimators for the relative risk. In Section~\ref{sec:proposed-estimators}, we propose methods for unbiased estimation, covariate adjustment, and handling partial interference.   In Section~\ref{sec:stepped-wedge}, we extend the set-up and results for parallel-arm CR-TNDs to stepped-wedge CR-TNDs. In Section~\ref{sec:simulation}, we support our theoretical results with simulation studies. In Section~\ref{sec: data-application}, we re-analyze  AWED study data providing a comparison of  estimators. In Section~\ref{sec:discussion}, we summarize our results and discuss future directions.

\section{Definition and assumptions}
\subsection{Notation}\label{subsec:setup}
Consider a cluster-randomized clinical trial that contains $m$ clusters. Each cluster $i$, $i \in\{ 1,\dots, m\}$, contains $n_i$ subjects defining the population of interest. For subject $j = 1,\dots, n_i$ in cluster $i$, let $Y_{ij}$ be a binary test-positive outcome, i.e., the outcome of interest, and $Z_{ij}$ be the binary test-negative outcome, which is used to assist the inference for the outcome of interest. In the AWED study, $Y_{ij}$ is an indicator of current dengue infection and $Z_{ij}$ is an indicator of other febrile illnesses (OFI). The value of $(Y_{ij}, Z_{ij})$ is determined by a single laboratory test, where a test-positive result indicates $(Y_{ij}, Z_{ij}) = (1, 0)$ and a test-negative result indicates $(Y_{ij}, Z_{ij}) = (0,1)$. 
If a subject has no disease, then $(Y_{ij}, Z_{ij}) = (0,0)$; however, a subject cannot have $Y_{ij} = Z_{ij} = 1$, e.g., OFI means febrile illnesses other than dengue in the AWED study.

In the context of TND, subjects may have different healthcare-seeking behavior, which impacts the observed data. For each participant, we define $S_{ij}$ as an indicator of whether the participant $j$ in cluster $i$ would seek healthcare if they had symptoms that are related to the outcome of interest and/or OFIs.
For example, in the AWED study, $S_{ij} = 1$ means that the participant would go to a clinic if they had an acute fever. We note that $S_{ij}$ is different from a non-missing indicator for $Y_{ij}$ and $Z_{ij}$. If a participant had $S_{ij} =1$ and $Y_{ij} = Z_{ij}=0$, then they would not seek healthcare (since they had no relevant symptoms) and $Y_{ij}, Z_{ij}$ would be not observed, i.e., the data will never show $Y_{ij} = Z_{ij}=0$; if a participant had $S_{ij} =1$ and $\max\{Y_{ij}, Z_{ij}\}=1$, then we would observe both $Y_{ij}$ and $Z_{ij}$ since they were ascertained by the same laboratory test.


In a cluster-randomized trial, let $A_i$ be the intervention indicator for cluster $i$ with $A_i =1$ if assigned to the intervention group and $0$ otherwise. We use the Neyman-Rubin potential framework, which makes the consistency assumption:
\begin{align*}
    Y_{ij} &= Y_{ij}(A_i) = A_i Y_{ij}(1) + (1-A_i) Y_{ij}(0), \\
    Z_{ij} &= Z_{ij}(A_i) = A_i Z_{ij}(1) + (1-A_i) Z_{ij}(0),
\end{align*}
where $Y_{ij}(a)$ and $Z_{ij}(a)$ are potential outcomes if the cluster $i$ were assigned intervention ($a=1$) or control ($a=0$). In addition, the potential outcomes are defined by the cluster-level hypothetical intervention, that is, all subjects in a cluster take the same intervention, avoiding the issue of handling  intracluster interference.

Similarly to the consistency assumption above, we assume $$S_{ij} = S_{ij}(A_i) = A_i S_{ij}(1) + (1-A_i) S_{ij}(0),$$ where $S_{ij}(a)$ encodes the healthcare-seeking behavior of participant $j$ if cluster $i$ were assigned intervention ($a=1$) or control ($a=0$). 
Our set-up for healthcare-seeking behavior is different from the literature on TNDs for observational studies \citep{sullivan2016theoretical, westreich2016invited, chua2020use}, which regarded $S_{ij}$ as a pre-intervention confounder.
For a cluster-randomized trial, we treat healthcare-seeking behavior as a post-randomization variable and allow the intervention to change such behavior, i.e. $S_{ij}(1) \ne S_{ij}(0)$.
This could happen when intervention is unblinded, and knowledge of intervention allocation may change  healthcare-seeking behavior.

As in \cite{jewell2019analysis}, we adopt the randomization inference framework and assume that all counterfactuals, $\{Y_{ij}(a), Z_{ij}(a), S_{ij}(a):a = 0,1,\ i = 1,\dots,m,\ j = 1,\dots, n_i\}$, are fixed numbers instead of random variables. We assume $m_1 = \sum_{i=1}^m A_i$ is fixed and
\begin{equation}\label{eq:intervention-allocation}
    P\{(A_1,\dots, A_m) = (a_1,\dots, a_m)\} = \frac{1}{\binom{m}{m_1}}
\end{equation}
for all $(a_1,\dots, a_m)$ that satisfy $\sum_{i=1}^m a_i = m_1$ and $a_i \in \{0,1\}$ for $i = 1,\dots, m$. Then the only randomness in $(Y_{ij}, Z_{ij}, S_{ij})$ comes from intervention allocation.

In a CR-TND, the observed data are defined as
\begin{equation*}
   \left(A_i, O_i^Y, O_i^Z\right), \quad i = 1,\dots, m, 
\end{equation*}
where, for each cluster $i$, $O_i^Y = \sum_{j=1}^{n_i} S_{ij}Y_{ij}$ is the observed test-positive counts and $O_i^Z = \sum_{j=1}^{n_i} S_{ij}Z_{ij}$ is the observed test-negative counts.
We, however,  do not observe the total case counts for either outcome (i.e., $\sum_{j=1}^{n_i}Y_{ij}$ and $\sum_{j=1}^{n_i}Z_{ij}$) or the number of healthcare seekers (i.e., $\sum_{j=1}^{n_i}S_{ij}$).

In the AWED study, $O_i^Y$ is the number of dengue cases among healthcare seekers, and $O_i^Z$ is the number of OFI cases among healthcare seekers. Due to the mechanism of the passive surveillance system, no information is recorded for those people with dengue or OFI who choose not to seek healthcare. Furthermore, the number of healthcare seekers is not known either, since we do not observe potential healthcare seekers who have no disease. 

The parameter of interest is the relative risk on the outcome of interest comparing intervention versus control, defined as
\begin{equation*}
    \lambda = \frac{\sum_{j=1}^{n_i} Y_{ij}(1) }{\sum_{j=1}^{n_i} Y_{ij}(0)},
\end{equation*}
where we assume that $\sum_{i=1}^{n_i} Y_{ij}(0) > 0$ and the relative risk $\lambda$ is constant across clusters. We discuss in Section \ref{subsec: unbiased-estimator} how to test and relax the latter  assumption  by adapting existing methods. Our goal is to make inference on $\lambda$ based on the observed data.

\subsection{Assumptions for CR-TNDs}
We make the following assumptions for estimating $\lambda$, which we refer to as the assumptions for a CR-TND. These assumptions are also seen in \cite{haber2015probability, anders2018cluster,dufault2020analysis}, where they are described under the super-population framework (i.e., assuming the data from each cluster $i$ are independent, identically distributed samples from a common distribution). We modify these assumptions to fit into the randomization inference framework as follows, which places no additional assumption on the counterfactual distributions.





\vspace{10pt}
\noindent\textbf{Assumption 1. (CR-TNDs)} For each $i = 1,\dots, m$,

(i) Intervention has no effect on the test-negative outcomes: $\sum_{j=1}^{n_i}Z_{ij}(1) =\sum_{j=1}^{n_i} Z_{ij}(0)$.

(ii) The relative proportion of healthcare seekers across intervention and control arms does not differ between test-positives and test-negatives:
\begin{equation*}
    \frac{p_i(1,Y)}{p_i(0,Y)} = \frac{p_i(1,Z)}{p_i(0,Z)},
\end{equation*}
where, for $a = 0,1$, $p_i(a,Y) = \frac{\sum_{j=1}^{n_i} S_{ij}(a) Y_{ij}(a)}{\sum_{j=1}^{n_i} Y_{ij}(a)}$ and $p_i(a,Z) = \frac{\sum_{j=1}^{n_i} S_{ij}(a) Z_{ij}(a)}{\sum_{j=1}^{n_i} Z_{ij}(a)}$ are the proportion of healthcare seekers among subjects with $Y_{ij}(a)=1$ and $Z_{ij}(a)=1$, respectively. 


\vspace{10pt}
Assumption 1 (i) indicates that $Z_{ij}$ is a negative-control outcome, and this assumption is commonly made for test-negative designs. For Assumption 1 (ii), we assume that if the intervention has an impact on healthcare-seeking behaviors, then such an impact must be the same in test-negative and test-positive populations. 
We denote $\alpha_i = \frac{p_i(1,Y)}{p_i(0,Y)} = \frac{p_i(1,Z)}{p_i(0,Z)}$, which represents the relative change of ascertainment probability, i.e., the probability of being tested, comparing intervention and control arms in the test-positive population (those with $Y_{ij}(a) =1$) or the test-negative population (those with $Z_{ij}(a) =1$). 

Assumption 1 characterizes the requirements for a CR-TND. Assumption 1 (i) requires that the intervention is not associated with test-negative illnesses. In the AWED study, this assumption is plausible since \textit{Wolbachia}-infected mosquito deployments have no effect on controlling OFI (so long as the latter does not include other flaviviruses such as Zika or chikungunya). For studies of influenza  vaccine effectiveness, this assumption, however, might be debatable as argued by \cite{haber2015probability}. Assumption 1 (ii) suggests that an unblinded intervention assignment should have the same impact on healthcare-seeking behavior among the test-positive population and test-negative population. This assumption is usually satisfied if the test-negative disease has similar symptoms compared to the disease of interest so that a patient cannot self-diagnose based on symptoms. 
For example, patients with dengue or OFI have common symptoms including fever and rash, and, hence, the true disease status (dengue or OFI) is ``blinded'' to patients before going to clinics. If the knowledge of intervention allocation changes the healthcare-seeking behavior of a patient, then such changes are not related to the true disease status due to a lack of knowledge of test status. This assumption is likely to hold in the AWED study as participants only know their test status days after seeking care. As a result, $\alpha_i$ is identical for test-positives and test-negatives. On the contrary, when the reasons for seeking healthcare differentiate between test-positives and test-negatives, then Assumption 1 (ii) may not hold since the knowledge of intervention allocation may have different impacts on  healthcare-seeking behavior.

An alternative set of assumptions made for test-negative designs were provided by \cite{jackson2013test}, and \cite{jewell2019analysis}, which are similar, but differ from Assumption 1. We describe these assumptions as Assumption 1' below.

\vspace{10pt}
\noindent\textbf{Assumption 1'.} For each $i = 1,\dots, m$,

(i') Among healthcare seekers, the incidence of test-negative outcomes does not differ between intervention and control:
\begin{align*}
    \frac{\sum_{j=1}^{n_i} S_{ij}(1)Z_{ij}(1)}{\sum_{j=1}^{n_i} S_{ij}(1)} = \frac{\sum_{j=1}^{n_i} S_{ij}(0)Z_{ij}(0)}{\sum_{j=1}^{n_i} S_{ij}(0)}.
\end{align*}

(ii') The intervention effect among healthcare seekers is generalizable to the whole population:
\begin{align*}
\lambda = \frac{\sum_{j=1}^{n_i} Y_{ij}(1)}{\sum_{j=1}^{n_i} Y_{ij}(0)} = 
    \frac{\sum_{j=1}^{n_i} S_{ij}(1)Y_{ij}(1)}{\sum_{j=1}^{n_i} S_{ij}(1)} \big/ \frac{\sum_{j=1}^{n_i} S_{ij}(0)Y_{ij}(0)}{\sum_{j=1}^{n_i} S_{ij}(0)}.
\end{align*}

\vspace{10pt}
Compared to Assumption 1 (i), Assumption 1' (i') is similar, but made on the population of healthcare seekers instead of the entire population. Assumption 1' (ii') assumes external validity, which may be debatable in practice \citep{sullivan2016theoretical, westreich2016invited}. 
Similar to Assumption 1 (ii), Assumption 1' (i') implicitly requires that the population of test-positives and test-negatives have similar reasons for seeking healthcare. 
This emphasizes the importance of the test-negative conditions exhibiting similar symptoms to the test-positive condition. For example, consider a cluster-level intervention that has no effect on dengue with $S_{ij}(a) =1$ if one seeks care for either dengue or breaks a leg (two {\em dissimilar} conditions), and $Z_{ij}(a) =0$ for a dengue case and $Z_{ij}(a) =1$ for a broken leg.
For participants with broken legs, an (unblinded) dengue intervention would not change their healthcare-seeking behavior (i.e., $S_{ij}(1) = S_{ij}(0)$)  or their test results (i.e, $Z_{ij}(1) = Z_{ij}(0)$), yielding $\sum_{j=1}^{n_i} S_{ij}(1)Z_{ij}(1) = \sum_{j=1}^{n_i} S_{ij}(0)Z_{ij}(0)$; for participants with incident dengue, however, the knowledge of intervention status may change their healthcare-seeking behavior, resulting in $\sum_{j=1}^{n_i} S_{ij}(1) \ne \sum_{j=1}^{n_i} S_{ij}(0)$ leading to a violation Assumption~1'~(i').

Despite the differences between Assumption 1 and Assumption 1', the  methods proposed in Section~\ref{sec:proposed-estimators} work under both sets of assumptions, allowing flexibility in trial planning. 
Assumption 1 and Assumption~1' both imply that, for each $i=1,\dots,m$,
\begin{equation}\label{eq:OYZ}
     O_i^Y(1) = \lambda c_i  O_i^Y(0), \quad O_i^Z(1) =c_i  O_i^Z(0),
\end{equation}
where, for $a = 0,1$, $O_i^Y(a) = \sum_{j=1}^{n_i} S_{ij}(a) Y_{ij}(a)$, and $O_i^Z(a) = \sum_{j=1}^{n_i} S_{ij}(a) Z_{ij}(a)$ are the potential observed data given intervention $a$, and $c_i$ is a quantity representing the relative ascertainment, i.e., differential healthcare-seeking behavior between intervention and control.
Under Assumption 1, $c_i = \alpha_i$ is the relative ascertainment in the population of test-positives and test-negatives; given Assumption 1', $c_i = \sum_{j=1}^{n_i} S_{ij}(1)/\sum_{j=1}^{n_i} S_{ij}(0)$ is the relative ascertainment in the whole cluster population, including  test-positives, test-negatives and people with no outcome-related symptoms.
For conciseness, we refer to $c_i$ as the ``relative ascertainment'' for cluster $i$ throughout.
Our results for unbiased estimation rely on Equation~(\ref{eq:OYZ}).


\section{Review of existing estimators for the relative risk}\label{sec:existing-estimators}
\subsection{The odds ratio estimator}\label{subsec:odds ratio-est}
The odds ratio estimator by \cite{jackson2013test} for $\lambda$ is defined as
\begin{equation}
    \widehat{\lambda} = \frac{\sum_{i=1}^m A_i O_i^Y}{\sum_{i=1}^m (1-A_i) O_i^Y} \frac{\sum_{i=1}^m (1-A_i) O_i^Z}{\sum_{i=1}^m A_i O_i^Z}.
\end{equation}
Assuming Assumption 1', $m = 2m_1$, and $c_i \equiv c$ for $i = 1,\dots m$, \cite{jewell2019analysis} showed that $E[\log(\widehat{\lambda})] \approx \log(\lambda)$ and derived an approximate variance formula for $\log(\widehat{\lambda})$, both under the permutation, or randomization, distribution. When the number of clusters is large, one can perform hypothesis testing and construct confidence intervals for $\lambda$ based on a Normal distribution, for which the validity is guaranteed by the Central Limit Theorem and Delta method. Of course, the null hypothesis of no intervention effect can be examined directly from the permutation distribution without approximation.

An assumption for the above procedure is that the relative ascertainment is the same across clusters, i.e., $c_i \equiv c$. 
Without this assumption,  $\log(\widehat{\lambda})$ can be biased for $\log({\lambda})$, where the bias, given Equation~(\ref{eq:OYZ}), is
\begin{displaymath}
E[\log(\widehat{\lambda})] - \log({\lambda}) = E\left[\log\left\{\frac{\sum_{i=1}^m c_iA_i O_i^Y(0)}{\sum_{i=1}^m (1-A_i) O_i^Y(0)} \frac{\sum_{i=1}^m (1-A_i) O_i^Z(0)}{\sum_{i=1}^m c_iA_i O_i^Z(0)}\right\}\right].
\end{displaymath}
This bias depends on the relationship between $c_i, O_i^Y(0), O_i^Z(0)$ and can be large if $c_i$ is strongly correlated with $O_i^Y(0)/O_i^Z(0)$, for example when relative ascertainment of participants tends to be greater among clusters with a higher (or lower) frequency of test-positives than test-negatives under no intervention. 
In Section~\ref{sec:simulation}, we use an AWED-based simulation to demonstrate that such bias can be potentially large and result in an inflated type I error.

The assumption, $c_i \equiv c$, can be relaxed if we alternatively assume that the data from each cluster are identically distributed and perform asymptotic results,  typically used in a super-population framework.
This alternative assumption, is debatable, however, when the number of clusters is limited or when the population of interest is limited entirely to the observed clusters. Under the randomization inference framework, we aim here to estimate $\lambda$ without assuming $c_i \equiv c$.

\subsection{The test-positive fraction estimator}\label{subsec:test-positive-fraction}
\cite{jewell2019analysis} proposed a test-positive fraction estimator for $\lambda$, based on the following statistic:
\begin{align*}
    T = \frac{1}{m_1} \sum_{i=1}^m  \frac{A_iO_i^Y}{O_i^Y + O_i^Z} - \frac{1}{m-m_1} \sum_{i=1}^m  \frac{(1-A_i)O_i^Y}{O_i^Y + O_i^Z}.
\end{align*}
Conditioning on the observed test-positive and test-negative counts, i.e., $O_{+}^Y = \sum_{i=1}^m O_i^Y$ and $O_{+}^Z=\sum_{i=1}^m O_i^Z$, and assuming $m_1 = m/2$, the conditional expectation of $T$, i.e., $E[T|O_{+}^Y, O_{+}^Z]$, is approximated by $\widetilde{E}_T$, defined as
\begin{equation}\label{eq: ET}
 \widetilde{E}_T= \frac{2r(\lambda^2-1)}{\{(2+r)\lambda+r\}\{r\lambda + 2 + r\}}, 
\end{equation}
where $r = O_{+}^Z/O_{+}^Y$; in addition, an approximate (permutation) variance estimator for $T$ was also derived. A point estimator of $\lambda$ is obtained by solving Equation~(\ref{eq: ET}) with $\widetilde{E}_T$ substituted by the observed $T$. 
Conditioning on $O_{+}^Y$ and $O_{+}^Z$, the test-positive fraction estimator can be used for hypothesis testing (using the permutation distribution) and constructing confidence intervals, while the variance for the estimator is less conveniently obtained.

It is reasonable to assume that $O_{+}^Y$ and $O_{+}^Z$ are fixed quantities if clusters are similar such that a different intervention allocation has no impact on the observed test-positive or test-negative counts. In general, however, intervention allocation will generally cause variation in observed test-positive and test-negative counts.
Specifically, Equation~(\ref{eq:OYZ}) implies that $O_{+}^Y = \sum_{i=1}^m A_i O_i^Y(1) + \sum_{i=1}^m (1-A_i) O_i^Y(0) = \sum_{i=1}^m (\lambda c_i A_i + 1-A_i)O_i^Y(0) $, which is a fixed quantity only if $(\lambda c_i-1)O_i^Y(0)$ is a constant across $i$. As a result, the approximate of $E[T|O_{+}^Y, O_{+}^Z]$ in Equation~(\ref{eq: ET}) is also random since it involves $r$, the ratio of observed test-negatives and test-positives. In our setting, we can derive
\begin{equation}\label{eq:ET-ET}
    E\left[T - \widetilde{E}_T\right] = \frac{1}{m}\sum_{i=1}^m \frac{(\lambda-1)U_i}{(\lambda + U_i)(1 + U_i)} - E\left[\frac{2r(\lambda^2-1)}{\{(2+r)\lambda+r\}\{r\lambda + 2 + r\}}\right],
\end{equation}
where $U_i =  O_i^Z(0)/ O_i^Y(0), i= 1,\dots, m$ are fixed quantities, and the expectation is over the permutation distribution without conditioning on $O_{+}^Y$ and $O_{+}^Z$. In the special case that $\lambda = 1$, Equation~(\ref{eq:ET-ET}) yields that $E\left[T - \widetilde{E}_T\right] = 0$ and the above inference based on $T$ remains valid; in other cases, $E\left[T - \widetilde{E}_T\right]$ may be non-zero which can lead to bias and inflated type I error. In Section~\ref{sec:simulation}, our simulation study shows that the bias can be either large or small depending on  $\lambda$.

\subsection{Model-based estimators}\label{subsec: model-based-methods}
For analysis of cluster-randomized trials, two commonly-used models are generalized linear mixed models (GLMM, \citealp{breslow1993approximate}) and generalized estimating equations (GEE, \citealp{liang1986longitudinal}). In CR-TNDs, we consider logistic regression for both models, since the observed outcomes naturally produce binary data with each cluster containing $O_i^Y$ ones and $O_i^Z$ zeros. For GLMM, we include an intercept and main terms for intervention assignment and cluster-level covariates as fixed effects and a cluster-level random intercept; for GEE, we consider the same fixed effects and an exchangeable within-cluster covariance structure. Both models estimate $\log(\lambda)$ by the regression coefficient of the intervention assignment fixed effect. 

The performance of GLMM and GEE for analyzing cluster-randomized trials has been extensively studied \citep{murray2004design, mcneish2016modeling, jewell2019analysis}. GEE targets a marginal effect as we consider here, while GLMM estimates a cluster-specific effect, which can be different from the marginal estimand of interest. Both models rely on strong model assumptions that are challenging to verify, such as the generalized linear model and the correlation structure. When these assumptions are violated,  the relative effect estimation and its variance estimation might be not valid. In addition, GEE is also known for suffering from inflated type I error when the number of clusters is small. Further, when the relative ascertainment $c_i$ varies across clusters in a CR-TND, both methods can be biased and we demonstrate this directly in the simulation study.  


\section{Proposed randomization-inference methods for CR-TNDs} \label{sec:proposed-estimators}
\subsection{The unbiased log-contrast estimator}\label{subsec: unbiased-estimator}
We define $L_{i} = \log(O_i^Y) - \log(O_i^Z)$, which is the log-contrast between test-positives and test-negatives among healthcare seekers. Using the notation of $O_i^Y(a)$ and $O_i^Z(a)$ defined in Equation~(\ref{eq:OYZ}), we further define $L_{i}(a) = \log\{O_i^Y(a)\} - \log\{O_i^Z(a)\}$ as the potential log-contrast given intervention $a \in \{0,1\}$. The consistency assumption implies that $L_i  = A_i L_i(1) + (1-A_i)L_i(0)$. In addition, given Equation~(\ref{eq:OYZ}), we have $L_i(1) = \log(\lambda) + L_i(0)$ for $i = 1,\dots,m$, a constant intervention effect model. Note that $L_i$ is a simple log odds transformation of the test-positive fraction of Section~\ref{subsec:test-positive-fraction}.

The log-contrast estimator is defined as
\begin{equation}
    \widehat{\log(\lambda)} = \frac{1}{m_1} \sum_{i=1}^m A_iL_i - \frac{1}{m - m_1} \sum_{i=1}^m (1-A_i)L_i,
\end{equation}
which is unbiased for $\log(\lambda)$. The variance of $\widehat{\log(\lambda)}$ is 
\begin{align*}
    Var\left(\widehat{\log(\lambda)}\right) = \frac{m}{m_1(m-m_1)} \frac{1}{m-1} \sum_{i=1}^m\{L_i(0) - \overline{L(0)}\}^2,
\end{align*}
where $\overline{L(0)} = m^{-1} \sum_{i=1}^m L_i(0)$, and can be unbiasedly estimated by 
\begin{align*}
    \widehat{Var}\left(\widehat{\log(\lambda)}\right) = \frac{1}{m_1} \widehat{\sigma}_1^2  + \frac{1}{m-m_1} \widehat{\sigma}_0^2,
\end{align*}
where $\widehat{\sigma}_a^2$ for $a = 0,1$ is the sample variance of $L_i$ with $A_i = a$. Given the unbiased variance estimator, one can construct the statistic $T= \widehat{\log(\lambda)}/\sqrt{\widehat{Var}\left(\widehat{\log(\lambda)}\right)}$ and construct the confidence interval for $\log(\lambda)$ and $\lambda$ based on the Normal distribution. Of course, an exact test of the null hypothesis is available using the permutation distribution directly.  
Compared to the estimators defined in Section~\ref{sec:existing-estimators}, our proposed estimator is able to eliminate bias when the relative ascertainment varies across clusters. In addition, the log-contrast estimator is also able to handle unequal randomization (i.e.,  $m\ne 2m_1$).

The above tests are based on the assumption that the relative risk is constant across clusters, i.e., $\lambda_i \equiv \lambda$. This assumption can be tested following the method of \cite{ding2016randomization} on $(L_1,\dots, L_m)$. When we reject the null hypothesis that $\lambda_i \equiv \lambda$, we can still estimate the average intervention effect $\tau = \frac{1}{m}\sum_{i=1}^m\log(\lambda_i)$ by the above test-statistic $T$, while the variance estimator $\widehat{Var}\left(\widehat{\log(\lambda)}\right)$ overestimates the true variance \citep{aronow2014sharp}, indicating that inference is still valid but can be conservative.

\subsection{Covariate adjustment}

Given the constant intervention effect model on $L_i$, we are able to use the covariate-adjusted estimator of \cite{lin2013agnostic} and \cite{li2017general} to improve precision. For each cluster $i$, let $\bX_i = (X_{i1}, \dots, X_{ip})$ be a $p$-dimensional vector of cluster-level baseline variables. We assume that each $\bX_i$ is a fixed vector.

The covariate-adjusted estimator for $\log(\lambda)$ is defined as
\begin{equation}\label{eq:cov-adj}
  \widehat{\log(\lambda)}_{\bbeta}  = \widehat{\log(\lambda)}  -  \bbeta^\top\left\{\frac{1}{m_1} \sum_{i=1}^m A_i\bX_i - \frac{1}{m - m_1} \sum_{i=1}^m (1-A_i)\bX_i\right\},
\end{equation}
where $\bbeta \in \mathbb{R}^p$ can be any constant vector. Direct calculation shows that $ \widehat{\log(\lambda)}_{\bbeta}$ is unbiased for $\log(\lambda)$.

Given the assumption of a constant intervention effect, the variance of $\widehat{\log(\lambda)}_{\bbeta}$ is minimized at $\bbeta^* = V(\bX)^{-1} C(\bX, L(0)),$
where $V(\bX) = \frac{1}{m-1} \sum_{i=1}^m (\bX_i - \overline{\bX})(\bX_i - \overline{\bX})^\top$ and $C(\bX, L(0)) = \frac{1}{m-1} \sum_{i=1}^m (\bX_i - \overline{\bX})(L_i(0) - \overline{L(0)})^\top$ with $\overline{\bX} = \frac{1}{m}\sum_{i=1}^m \bX_i$. 
According to \cite{roth2021efficient},  $Var(\widehat{\log(\lambda)}) = Var(\widehat{\log(\lambda)}_{\bbeta^*}) + \frac{m}{m_1(m-m_1)}\bbeta^*{}^\top V(\bX)\bbeta^*$, quantifying the variance reduction through covariate adjustment. In practice, since the values of $L_i(0)$ are not fully observed, $\bbeta^*$ is unknown. However, one can construct an estimator for $\bbeta^*$ by $\widehat{\bbeta} = \frac{m_1}{m} \widehat{\bbeta}_1 + \frac{m-m_1}{m} \widehat{\bbeta}_0$, where $\widehat{\bbeta}_1$ and $\widehat{\bbeta}_0$ are the sample least-square coefficients of $L_i$ on $\bX_i$ for the intervention group and control group, respectively. Asymptotically, $\widehat{\bbeta}$ is equivalent to $\bbeta^*$ in terms of efficiency \citep{li2017general, roth2021efficient}. The variance of $ \widehat{\log(\lambda)}_{\widehat{\bbeta}}$ can be estimated by
\begin{displaymath}
\frac{1}{m_1} \widehat{\sigma}_{1,\widehat{\bbeta}}^2  + \frac{1}{m-m_1} \widehat{\sigma}_{0,\widehat{\bbeta}}^2,
\end{displaymath}
where $\widehat{\sigma}_{1,\widehat{\bbeta}}^2$ and $\widehat{\sigma}_{1,\widehat{\bbeta}}^2$ are the unbiased residual variance estimator for the intervention group and control group after regressing $L_i$ on covariates $\bX_i$, respectively.

\subsection{Dose-response relationship under partial compliance}\label{sec: dose-response}

In cluster-randomized clinical trials, non-compliance or partial compliance may occur after intervention is assigned. For example, in the AWED study, a \textit{Wolbachia} exposure index (WEI) was used to measure an individual-level compliance status, defined as a weighted score accounting for the mobility of participants immediately prior to symptom onset and the percentage of \textit{Wolbachia}-infected mosquitoes \citep{utarini2021efficacy} in all city locations as measured through mosquito trapping. The variable, WEI, is a continuous variable taking values in $[0,1]$ with a larger value indicating more exposure to the intervention (e.g., 1 means no mobility and 100\% \textit{Wolbachia}-infected mosquitoes around the place of residence).
The cluster-level WEI is then defined as the average of WEI among enrolled patients within a cluster, representing the cluster-level compliance status.
Averaged over the follow-up period of the AWED study, the cluster-level WEI ranges from 0.66 to 0.75 in intervention clusters and  0.22 to 0.44  in  control clusters. 
We now focus on quantifying the intervention effect based on the actual intervention ``received".

We consider an instrumental variable method with a linear model to capture the dose-response relationship for a general CR-TND. 
For each cluster $i$, define $D_i \in [0,1]$ as the cluster-level measure of intervention received, and  $L_i(d,a)$ as the potential outcome of $L_i$, the log-contrast, if cluster $i$ received intervention $a, a \in\{0,1\}$ and complied with the intervention as measured by $d , d\in [0,1]$. Here $L_i(d,a)$ extends $L_i(a)$ defined in Section~\ref{subsec: unbiased-estimator} by accounting for compliance:  if $d = a$, then the cluster $i$ receives a perfect exposure to the intervention, or control, respectively as $a=1$ or $a=0$, and $L_i(a,a) = L_i(a)$; if $d\in(0,1)$, then the cluster $i$ only has partial intervention compliance. We make the following assumptions to allow identification of the dose-response relationship.

\vspace{10pt}
\noindent\textbf{Assumption 2. (Linear Dose-response relationship)} For $i=1,\dots, m$,

(i) Consistency: $L_i = L_i(D_i,A_i)$.

(ii) Exclusion restriction: $L_i(d,a) = L_i(d)$ for any $a \in\{0,1\}$ and $d\in[0,1]$. 

(iii) The linear model: $L_i(d) = L_i(0) + \beta d$ for $\beta \in \mathbb{R}$.

\vspace{10pt}
Assumption 2 (i) connects the observed data and the counterfactual outcome by letting $d = D_i$ and $a = A_i$. 
Assumption 2 (ii) indicates that
the intervention assignment has no direct effect on the outcomes given the actual intervention score $d$. Both Assumptions 2 (i) and (ii) are standard assumptions for instrumental variable methods \citep{angrist1996identification}. 
Assumption 2 (iii) specifies a linear model, indicating that the intervention effect is proportional to ``dose" received. 

We base our approach on randomization inference with instrumental variables \citep[Section 5.4]{rosenbaum2002observational} and its extension for group randomization \citep{small2008randomization}, generalizing from a binary intervention setting to a continuous intervention setting. 
In the special case of perfect compliance, i.e. $D_i = A_i$ for all $i$, Assumption 2 is compatible with Assumption 1 or 1' with $\beta = \log\lambda$.

Given the null hypothesis $H_0: \beta = \beta_0$, the covariate-adjusted estimator can be computed and its variance  estimated similarly by setting $L_i(0) = L_i - \beta_0 D_i$. Then hypothesis testing can be performed for $H_0$ based on A Normal approximation, and a confidence interval can be obtained by inverting such tests \citep[Section 2.6]{rosenbaum2002observational}.

\section{Extensions to the stepped-wedge design}\label{sec:stepped-wedge}
\subsection{Set-up}
In a stepped-wedge design, an intervention is sequentially  assigned to clusters such that all clusters start with no intervention, and, at the end, all receive the intervention.
Let $\{1,\dots, T\}$ be the time window of intervention allocation. For each $t \in \{1,\dots, T\}$, $q_t$ clusters start the intervention such that $\sum_{t=1}^T q_t = m$. Let $A_i$ be the time that cluster $i$ is assigned to the intervention group and $\Omega$ be the set of all possible values of $(A_1,\dots, A_m)$. Then $\Omega$ contains $\binom{m}{q_1,\dots,q_t}$ entries. The randomization scheme implies that $P\{(A_1,\dots, A_m)=(t_1,\dots, t_m)\} = \frac{1}{\binom{m}{q_1,\dots,q_t}}$ for any $(t_1,\dots, t_m) \in \Omega$.

Similar to the definition of $Y_{ij}(a), Z_{ij}(a), S_{ij}(a)$ in Section~\ref{subsec:setup}, we analogously define the counterfactuals $Y_{ijt}(a), Z_{ijt}(a), S_{ijt}(a)$, which represent the potential test-positive outcome, test-negative outcome, and healthcare-seeking behavior for participant $j$ in cluster $i$ at time $t$, respectively, if the cluster $i$ has intervention status $a$ at time $t$, $t \in \{1,\dots,T\}$ and $a \in \{0,1\}$. 
Here we make the simplifying assumption that the intervention effect is immediate and not altered by the duration of the intervention. We again make the following consistency assumption:
\begin{align*}
   Y_{ijt} &= I\{A_i \le t\} Y_{ijt}(1) + I\{A_i > t\} Y_{ijt}(0), \\
   Z_{ijt} &= I\{A_i \le t\} Z_{ijt}(1) + I\{A_i > t\} Z_{ijt}(0), \\
   S_{ijt} &= I\{A_i \le t\} S_{ijt}(1) + I\{A_i > t\} S_{ijt}(0).
\end{align*}
Then the observed data are $(A_i, \sum_{j=1}^{n_{it}} S_{ijt}Y_{ijt}, \sum_{j=1}^{n_{it}} S_{ijt}Z_{ijt})$ for $i=1,\dots, m$ and $t = 1,\dots, T$, where $n_{it}$ is the number of subjects in cluster $i$ at time $t$.

Our goal is to estimate the relative risk comparing intervention to control, defined as, 
\begin{equation*}
    \lambda = \frac{\sum_{j=1}^{n_i} Y_{ijt}(1)}{\sum_{j=1}^{n_i} Y_{ijt}(0)},
\end{equation*}
for each $i = 1,\dots,m$ and $t = 1,\dots, T$, where we assume that $\sum_{j=1}^{n_i} Y_{ijt}(0) >0$ and the relative risk is constant across clusters and time

We make the following assumption, an  extension of  Assumption 1 to stepped-wedge CR-TNDs (SW-TND):

\vspace{10pt}
\noindent\textbf{Assumption 3. (SW-TND)} For each $i=1,\dots, m$ and $t= 1,\dots, T$,

(i) The intervention has no effect on the test-negative outcomes: $\sum_{j=1}^{n_i}Z_{ijt}(1) =\sum_{j=1}^{n_i} Z_{ijt}(0)$.

(ii) At any time, the relative proportion of healthcare seekers  comparing intervention and control does not differ between test-positives and test-negatives:
\begin{equation*}
    \frac{p_{it}(1,Y)}{p_{it}(0,Y)} = \frac{p_{it}(1,Z)}{p_{it}(0,Z)} = c_{it},
\end{equation*}
where for $a = 0,1$, $p_{it}(a,Y) = \frac{\sum_{j=1}^{n_i} S_{ijt}(a) Y_{ijt}(a)}{\sum_{j=1}^{n_i} Y_{ijt}(a)}$ and $p_{it}(a,Z) = \frac{\sum_{j=1}^{n_i} S_{ijt}(a) Z_{ijt}(a)}{\sum_{j=1}^{n_i} Z_{ijt}(a)}$ are the proportion of healthcare seekers among subjects with $Y_{ijt}(a)=1$ and $Z_{ijt}(a)=1$, respectively.

\vspace{10pt}
For the stepped-wedge design, Assumption 3 essentially assumes Assumption 1 at each time point $t$. If the time period $T$ is relatively short, a test-negative design that satisfies Assumption 1 would be likely to imply Assumption 3 also. Note that we do not make any assumption on the temporal trend of the test-positive  or test-negative diseases, which can vary across clusters. Similar to the parallel-arm CR-TND, we use $c_{it}$ to denote the relative ascertainment for cluster $i$ at time $t$. 

\subsection{Estimation}
We define $L_{it} = \log\left\{\sum_{j=1}^{n_i} S_{ijt} Y_{ijt}\right\} - \log\left\{\sum_{j=1}^{n_i} S_{ijt} Z_{ijt}\right\}$ and, for $a = 0,1$, $L_{it}(a) = \log\left\{\sum_{j=1}^{n_i} S_{ijt}(a) Y_{ijt}(a)\right\} - \log\left\{\sum_{j=1}^{n_i} S_{ijt}(a) Z_{ijt}(a)\right\}$. With this set-up, and Assumption 3, it follows that
\begin{align*}
    L_{it} &= I\{A_i < t\} L_{it}(1) + I\{A_i \ge t\} L_{it}(0),\\
    L_{it}(1) &= \log(\lambda) + L_{it}(0).
\end{align*}
We construct the following estimator for $\log(\lambda)$, referred to as the stepped-wedge log-contrast estimator:  
\begin{equation}
    \widehat{\log(\lambda)}^{\textup{(SW)}} = \sum_{t=1}^{T-1} w_t\left\{ \frac{1}{m_t} \sum_{i=1}^m I\{A_i \le t\}L_{it} - \frac{1}{m - m_t} \sum_{i=1}^m I\{A_i > t\}L_{it} \right\},
\end{equation}
where $m_t = \sum_{t'=1}^t q_{t'}$ is the number of clusters in the intervention group at time $t$ and $w_t$ are arbitrary pre-specified weights with $\sum_{t=1}^{T-1}w_t = 1$. The proposed estimator $\widehat{\log(\lambda)}^{\textup{(SW)}}$ is a weighted average of the difference-in-means estimators across $t$. Since each of the difference-in-means estimator is unbiased for $\log(\lambda)$, then the estimator $\widehat{\log(\lambda)}^{\textup{(SW)}}$ is also unbiased for $\log(\lambda)$.

The variance of $\widehat{\log(\lambda)}^{\textup{(SW)}}$ is $\boldsymbol{w}^\top \mathbf{\Sigma} \boldsymbol{w}$, where
\begin{align*}
\boldsymbol{w} &= (w_1,\dots, w_{T-1})^\top, \\
    \mathbf{\Sigma} &=  \left(\begin{array}{ccc}
    \frac{m}{m_1(m-m_1)} S_{1,1}    &  \cdots & \frac{m}{m_{T-1}(m-m_1)} S_{T-1,1}\\
    \vdots     &  \ddots & \vdots \\
    \frac{m}{m_{T-1}(m-m_1)} S_{T-1,1} & \cdots & \frac{m}{m_{T-1}(m-m_{T-1})} S_{T-1,T-1}
    \end{array}\right), \\
    S_{t_1,t_2} &= \frac{1}{m-1} \sum_{i=1}^m \left\{L_{it_1}(0) - \overline{L_{t_1}(0)}\right\}\left\{L_{it_2}(0) - \overline{L_{t_2}(0)}\right\}, \\
    \overline{L_{t}(0)} &= \frac{1}{m}\sum_{i=1}^mL_{it}(0).
\end{align*} 
Given $\boldsymbol{w}$, the variance can be unbiasedly estimated by  substituting $\widehat{\mathbf{\Sigma}}$ for $\mathbf{\Sigma}$, where $\widehat{\mathbf{\Sigma}}$ is an unbiased estimator of ${\mathbf{\Sigma}}$ with the $(t_1,t_2)$-th entry $(t_1\le t_2)$ being $\frac{m}{m_{t_2-1}(m-m_{t_1})} \widehat{S}_{t_1,t_2}$, where
\begin{align*}
    \widehat{S}_{t_1,t_2} = \left\{\begin{array}{cc}
   \widehat{c}_{(1,1)}(t_1,t_2)      & \textrm{if } m_{t_1} \ge \max\{m_{t_2} - m_{t_1}, m - m_{t_2}\}, \\
   \widehat{c}_{(0,1)}(t_1,t_2)      &   \textrm{if } m_{t_2} - m_{t_1} \ge \max\{ m_{t_1}, m - m_{t_2}\}, \\
   \widehat{c}_{(0,0)}(t_1,t_2)      &  \textrm{if }  m - m_{t_2} \ge \max\{ m_{t_1}, m_{t_2} - m_{t_1}\},
    \end{array}\right.
\end{align*}
where $\widehat{c}_{(1,1)}(t_1,t_2)$,  $\widehat{c}_{(0,1)}(t_1,t_2)$, and  $\widehat{c}_{(0,0)}(t_1,t_2)$ are the sample covariance between $L_{it_1}$ and  $L_{it_2}$ among clusters with $A_i \le t_1$, $  t_1 < A_i \le t_2$, and $A_i > t_2$, respectively. 

For the weights $ \boldsymbol{w}$, a convenient option is to set $w_t = (T-1)^{-1}$. To optimize the efficiency of the proposed estimator, we can minimize the variance by setting 
\begin{equation}\label{eq:w*}
    \boldsymbol{w}^* = \frac{\mathbf{\Sigma}^{-1}\bone_{T-1}}{\bone_{T-1}^\top \mathbf{\Sigma}^{-1}\bone_{T-1}},
\end{equation}
where $\bone_{T-1}$ is a $(T-1)$-dimensional column vector with all entries 1.
Given the null hypothesis $H_0:\lambda = \lambda_0$, $\boldsymbol{w}^*$ can be exactly computed; in other cases, it can be approximated by plugging in $\widehat{\mathbf{\Sigma}}$. When $m$ is small and $T$ is large, however, such a plug-in estimator of $\boldsymbol{w}^*$ can be less stable than a pre-specified $\boldsymbol{w}$, since the inverse of $\widehat{\mathbf{\Sigma}}$ can be highly variable and potentially not invertible.



Given the proposed estimator and variance estimator,  statistical inference can then be performed as described in Section~\ref{subsec: unbiased-estimator}. Again, it is possible to carry out exact inference using the permutation distribution.

\section{Simulation studies}\label{sec:simulation}
\subsection{Simulation set-up}
We consider two simulation studies that are based on historical dengue and OFI data in Yogyakarta. The first simulation study verifies the unbiasedness of the proposed log-contrast estimator for CR-TNDs and compares it to the existing estimators defined in  Section~\ref{sec:existing-estimators}. The second simulation study focuses on the SW-TND setting, where we evaluate the performance of the stepped-wedge log-contrast estimator. For both simulations, we consider $\lambda = 1, 0.6, 0.2$, which represent a zero, moderate, and large intervention effect, respectively.

The first simulation study is based on reported (serious) dengue cases from 2013-2015, and OFI cases from 2014-2015 for each of the 24 clusters in Yogyakarta. In addition, cluster population size (measured in 10,000s) in the Year 2015 is used as a covariate. All data are available in Table S1 of the Supplementary Material of \cite{utarini2021efficacy}.
Since no information on relative ascertainment is available, we independently sample $c_i, i=1,\dots, 24$ from a Beta distribution with the shape parameters set as 0.5. Of note, relative ascertainment is generated once and applied to all simulated data sets, representing a fixed latent characteristic of the 24 clusters; otherwise, if $c_i$ is sampled for each simulated data set, then a distributional assumption is placed on relative ascertainment across data sets, a scenario not of our interest as discussed in Section~\ref{subsec: unbiased-estimator}.
From the above information,  $\widetilde{O}^Y_i, \widetilde{O}^Z_i, X_i$ denote ascertained dengue cases, OFI cases, and population size, respectively.

The data are simulated through the following steps.  First, we randomly generate \\ $\{O^Y_1(0),\dots, O^Y_{24}(0)\}$ from a multinomial distribution with parameters $(n^Y, \widetilde{O}^Y_1/n^Y, \dots, \widetilde{O}^Y_{24}/n^Y)$ and generate $\{O^Z_1(0),\dots, O^Z_{24}(0)\}$ from a multinomial distribution with parameters \\ $(n^Z, \widetilde{O}^Z_1/n^Z, \dots, \widetilde{O}^Z_{24}/n^Z)$, where $n^Y = \sum_{i=1}^{24} \widetilde{O}^Y_i$ and $n_Z = \sum_{i=1}^{24} \widetilde{O}^Z_i$. We then introduce correlation between the outcome and covariate by multiplying $O^Y_i(0)$ by $2X_i$ and dividing $O^Z_i(0)$ by $2X_i$. Next, we set $O^Y_i(1) = \lambda c_i O^Y_i(0)$ and $O^Z_i(1) =  c_i O^Z_i(0)$. Finally, letting $(A_1,\dots, A_{24})$ be the random intervention allocation following the distribution~(\ref{eq:intervention-allocation}) with $m_1 = 12$, we define $O^Y_i = A_iO^Y_i(1) + (1-A_i)O^Y_i(0)$ and $O^Z_i = A_iO^Z_i(1) + (1-A_i)O^Z_i(0)$. The simulated data are $\{(A_i, O_i^Y, O_i^Z, X_i)\}_{i=1}^{24}$.

For the second simulation study, we use observed dengue cases collected for every two consecutive years between 2003 and 2014 for each of the 24 clusters in Yogyakarta. Due to lack of data in the years 2004 and 2009, each cluster has nine data points. The data are available in the supplementary material of \cite[Table 2]{jewell2019analysis}. We use  $\widetilde{O}_{it}^Y$ to denote the above dengue cases for $i = 1,\dots, 24$ and $t = 1,\dots, 9$  and define $n_t^Y = \sum_{i=1}^{24} \widetilde{O}_{it}^Y$. 
OFI data are only available for the Year 2014-2015, which are the same as those used in first simulation study, and we keep the notation of $\widetilde{O}_{i}^Z$ and $n^Z$. To generate the OFI case counts for earlier years, we define $\widetilde{O}_{it}^Z = \widetilde{O}_{i}^Z n_t^Y/n_9^Y$ and $n_t^Z = n^Zn_t^Y/n_9^Y$. For each $t$, we generate $c_{it}$ and $O_{it}^Y(a), O_{it}^Z(a), a = 0,1$ following a similar procedure to  the first simulation study by substituting $(\widetilde{O}_{it}^Y, \widetilde{O}_{it}^Z)$ for $(\widetilde{O}_{i}^Y, \widetilde{O}_{i}^Z)$ except that the covariate $X_i$ is omitted. Specifically, $c_{it}$ is independently generated for each $i$ and $t$, representing both temporal and cluster variation of relative ascertainment. Intervention starts at $t=2$ and, for each $t = 2,\dots, 9$,  three untreated clusters are randomly selected to start intervention. We then define $O^Y_{it} = I\{A_i < t\} O^Y_{it}(1) +  I\{A_i \ge t\}O^Y_{it}(0)$ and $O^Z_{it} = I\{A_i < t\} O^Z_{it}(1) +  I\{A_i \ge t\}O^Z_{it}(0)$. The simulated data are $\{(A_i, O_{i1}^Y, \dots, O_{i9}^Y,O_{i1}^Z,\dots, O_{i9}^Z)\}_{i=1}^{24}$. Furthermore, our data generating distribution implies that, within each simulated data set, the temporal correlation of outcomes varies across clusters.

We simulate $10,000$ data sets and estimate the log relative risk, $\log(\lambda)$, for both simulation studies. In the first simulation, we compare the odds ratio estimator, test-positive fraction estimator, GLMM estimator, GEE estimator, and our proposed log-contrast estimator and covariate-adjusted estimator. For the test-positive fraction estimator, we follow the method of \cite{jewell2019analysis} and obtain $\hat{\lambda}$ by solving Equation~(\ref{eq: ET}) with plugged-in $T$ and $r$.

For the second simulation study, we compare  the stepped-wedge log-contrast estimator with equal weights or optimal weights to the GLMM and GEE estimators. 
To model temporal correlation, the GLMM includes random effects for the cluster intercept and the cluster-by-time intercept as suggested by \cite{ji2017randomization}; for GEE, we use an exchangeable correlation structure. Due to the limited number of clusters, the optimal weights $\boldsymbol{w}^*$ are estimated assuming the true relative risk is known, since otherwise, the estimated covariance matrix $\widehat{\boldsymbol\Sigma}$ is often not invertible; our simulation study for the optimal weights is hence  designed for hypothesis testing.

The comparison metrics are bias, standard error of the estimates, average of the standard error estimates, probability of rejecting the null hypothesis (i.e., $\lambda  = 1$), and the  coverage  probability  of  nominal  0.95  confidence  intervals based on Normal approximations.

\subsection{Simulation results}
Table~\ref{tab:simulation1} summarizes the results of the first simulation study. Among all estimators, our proposed log-contrast and covariate-adjusted estimators are unbiased and achieve the desired coverage probability, while all other estimators show bias for most values of $\lambda$. Due to varying relative ascertainment across clusters, the odds ratio, GLMM, and GEE estimators have bias and inflated type I error that results in 3-6\%, 4-8\%, and 8-9\% under-coverage, respectively. 
The test-positive fraction estimator is valid if $\lambda=1$, while its bias and type I error increase as the true relative risk moves away from the null; such a bias can be as high as 0.34 when there is a strong intervention effect. 
In terms of variance, since the covariate is designed to be prognostic, the covariate-adjusted estimator has 24\% smaller variance than the log-contrast estimator, showing that adjustment for prognostic baseline variables can improve precision. 
The model-based estimators have similar standard errors compared to the covariate-adjusted estimator, while their standard error estimates consistently underestimate the true standard error, i.e., ASE smaller than SE, which contributes to the inflated type I error of model-based inference. 
The test-positive fraction estimator has the smallest standard error among all estimators, whereas its validity is hampered by its bias; furthermore, a standard error estimate of the  test-positive fraction estimator is more difficult to compute directly and so is omitted here--see \cite{jewell2019analysis}. 

\begin{table}[htbp]
\caption{Simulation results under the CR-TND setting. For each of the estimators, we report the bias, standard error of the estimates (SE), average of the standard error estimates (ASE), probability of rejecting the null hypothesis (PoR), and the  coverage  probability  of  nominal  0.95  confidence  intervals based on Normal approximations (CP). The bias, SE, and ASE are on the log scale.}\label{tab:simulation1}
\centering
\begin{tabular}{crrrrrr}
  \hline
&Estimators & Bias & SE & ASE & PoR & CP \\ 
  \hline
\multirow{6}{*}{$\lambda=1~(\log\lambda=0)$}&Odds ratio & 0.14 & 0.32 & 0.31 & 0.09 & 0.91 \\ 
& Test-positive fraction & -0.00 & 0.26 & - & 0.06 & 0.94 \\ 
&  GLMM &  0.05 & 0.27 & 0.25 & 0.09 & 0.91 \\ 
& GEE & 0.14 & 0.26 & 0.24 & 0.14 & 0.86 \\ 
 &  Log-contrast & 0.00 & 0.31 & 0.32 & 0.06 & 0.94 \\ 
 &  Covariate-adjusted &  -0.00 & 0.27 & 0.27 & 0.06 & 0.94 \\ 
   \hline
\multirow{6}{*}{$\lambda=0.6~(\log\lambda=-0.51)$}&Odds ratio &0.14 & 0.32 & 0.31 & 0.20 & 0.89 \\ 
& Test-positive fraction & 0.08 & 0.26 & - & 0.37 & 0.94 \\ 
 & GLMM &  0.05 & 0.27 & 0.25 & 0.42 & 0.91 \\ 
 & GEE & 0.13 & 0.27 & 0.24 & 0.35 & 0.86 \\ 
 &  Log-contrast & 0.00 & 0.31 & 0.32 & 0.35 & 0.94 \\ 
&  Covariate-adjusted & -0.00 & 0.27 & 0.27 & 0.47 & 0.94 \\ 
\hline
\multirow{6}{*}{$\lambda=0.2~(\log\lambda=-1.61)$}&Odds ratio &0.14 & 0.32 & 0.35 & 1.00 & 0.92 \\ 
& Test-positive fraction &0.34 & 0.23 & - & 1.00 & 0.88 \\ 
 & GLMM &  0.07 & 0.27 & 0.25 & 1.00 & 0.89 \\ 
 & GEE & 0.11 & 0.28 & 0.25 & 1.00 & 0.87 \\ 
 &  Log-contrast & 0.00 & 0.31 & 0.32 & 1.00 & 0.94 \\ 
&  Covariate-adjusted & -0.00 & 0.27 & 0.27 & 1.00 & 0.94 \\
\hline
\end{tabular}
\end{table}

\begin{figure}[p]
    \centering
    \includegraphics[width=0.95\textwidth]{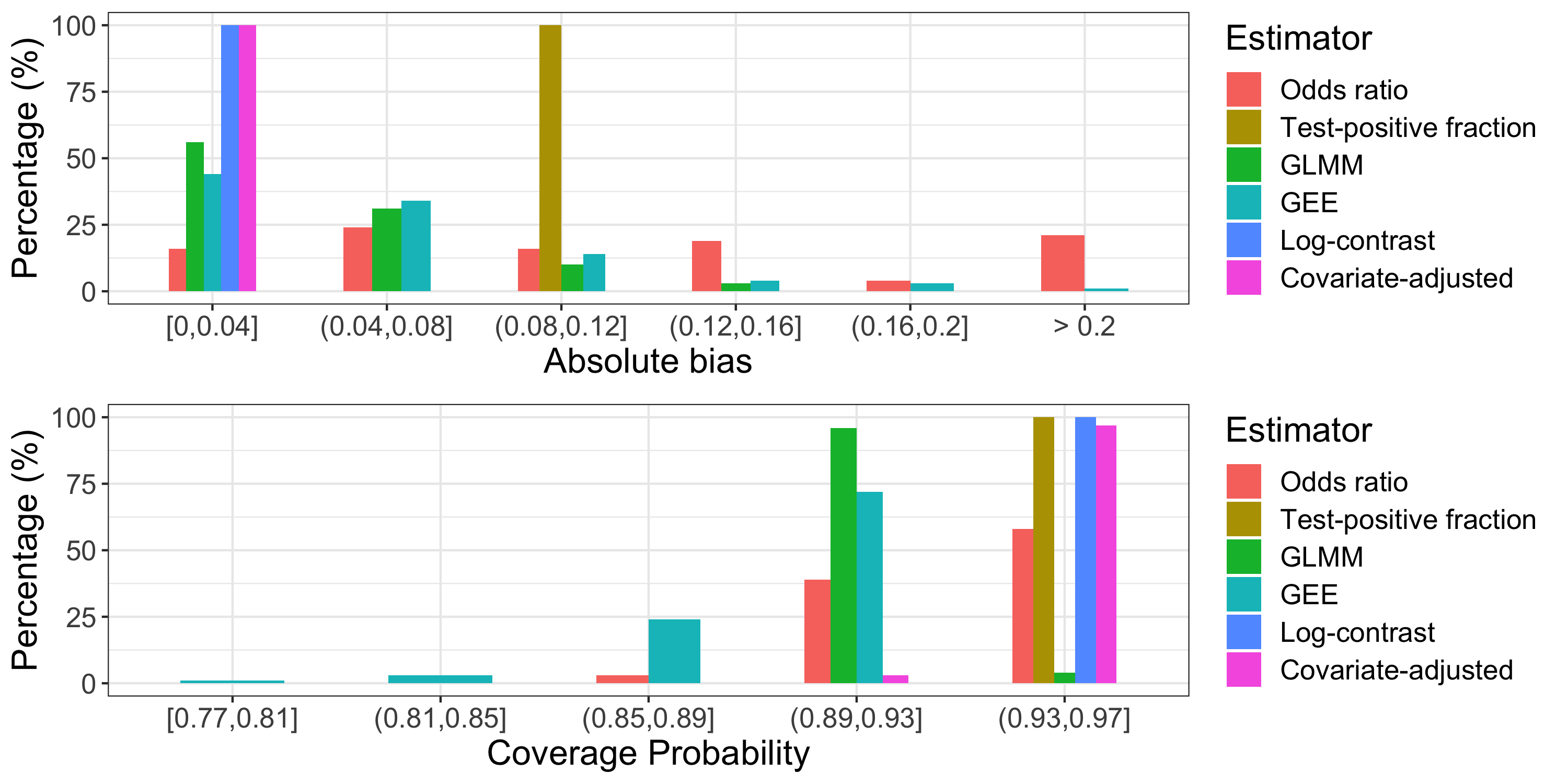}
    \caption{Distributions of the absolute value of bias (upper panel) and the coverage probability of Normal-approximated 95\% confidence interval (lower panel) for each estimator over 100 replications of the first simulation study. Each replication uses an independently-generated relative ascertainment, which causes variation in the bias and coverage probability.}\label{fig:fig1}
\end{figure}

To examine the impact of different choices of relative ascertainment, we independently repeat the first simulation study with $\lambda=0.6$ for 100 times, each with a distinct configuration of $c_i$'s. For each estimator, we present the distribution of bias  and coverage probability in Figure \ref{fig:fig1}, which further confirms the findings in Table~\ref{tab:simulation1}. Despite changes in relative ascertainment, the log-contrast and covariate-adjusted estimators maintain the property of unbiasedness and correct coverage probability. The other estimators, however, have more dispersed and shifted distributions of absolute bias and coverage probability, implying a bias and incorrect coverage probability, whose magnitude varies upon the characteristics of relative ascertainment.

Table~\ref{tab:stepped-wedge} displays the performance of estimators for the SW-TND. Similar to the first simulation study, GLMM and GEE lead to biased effect estimation, underestimation of the standard error, and under-coverage of the 95\% confidence interval. In contrast, the stepped-wedge log-contrast estimators remain unbiased and maintain the 0.05 type I error rate as desired. By using the optimal weights, the variance is reduced by 27\% compared to the equal weights. 
The equal weights, however, eliminate the uncertainty in the estimation of weights and, hence, have the advantage of being more stable for data with large $T$ and small $m$ over the optimal weights. Comparing model-based inference and randomization-based inference, we observe that the former is more powerful than the latter under a stepped-wedge design, resembling the results by Table 5 of \cite{ji2017randomization}. The power gain of model-based inference comes from additional model assumptions: both GLMM and GEE assume the temporal trend is the same across clusters, and directly model temporal correlation, while the randomization-based inference does not make such an assumption. When this assumption is violated, the validity of model-based inference is affected as reflected by the bias and inflated type I error in this simulation study.

\begin{table}[htbp]
\caption{Simulation results under the SW-TND setting. For the log-contrast estimator (with equal or optimal weight), GLMM estimator, and GEE estimator, we report its average bias (Bias), standard error estimates (SE), average of the standard error estimators (ESE), probability of rejecting the null hypothesis (PoR), and coverage probability (CP). The bias, SE, and ESE are on the log scale.}\label{tab:stepped-wedge}
\centering
\begin{tabular}{crrrrrr}
  \hline
&Estimators & Bias & SE & ESE & PoR & CP \\ 
  \hline
\multirow{4}{*}{$\lambda=1~(\log\lambda=0)$}  &  Log-contrast (equal weight) & 0.00 & 0.21 & 0.20 & 0.06 & 0.94 \\ 
& Log-contrast (optimal weight) & 0.00 & 0.18 & 0.18 & 0.05 & 0.95 \\ 
&  GLMM &  0.01 & 0.10 & 0.08 & 0.10 & 0.90 \\ 
& GEE & 0.06 &  0.06 & 0.07 & 0.11 & 0.89 \\ 
   \hline
\multirow{4}{*}{$\lambda=0.6~(\log\lambda=-0.51)$}  &  Log-contrast (equal weight) &  0.00 & 0.21 & 0.20 & 0.70 & 0.94 \\ 
& Log-contrast (optimal weight) & 0.00 & 0.18 & 0.18 & 0.80 & 0.95 \\ 
&  GLMM &  0.02 & 0.10 & 0.08 & 1.00 & 0.89 \\ 
& GEE & 0.06 & 0.06 & 0.07 & 1.00 & 0.87 \\ 
   \hline
\multirow{4}{*}{$\lambda=0.2~(\log\lambda=-1.61)$}  &  Log-contrast (equal weight) & 0.00 & 0.21 & 0.20 & 1.00 & 0.94 \\ 
& Log-contrast (optimal weight) & 0.00 & 0.18 & 0.18 & 1.00 & 0.95 \\ 
& GLMM &  0.03 & 0.11 & 0.10 & 1.00 & 0.91\\ 
& GEE & 0.07 & 0.06 & 0.08 & 1.00 & 0.89\\
   \hline
\end{tabular}
\end{table}


\section{Application to the AWED trial}\label{sec: data-application}
We re-analyze the AWED study using the estimators defined in Sections~\ref{sec:existing-estimators} and \ref{sec:proposed-estimators}. For each cluster, the dengue and OFI cases are aggregated across the follow-up period, respectively. For all estimators, a Normal approximation is used to construct confidence intervals. For GLMM, GEE, and covariate-adjusted estimators, the population size and the population proportion of children (age $<$ 15) are used as cluster-level baseline variables. The AWED study used covariate constrained randomization to achieve covariate balance and improve precision; this constrained randomization provided the basis for hypothesis testing but was not used in confidence interval calculations. For simplicity and illustration, we also assumed complete randomization in our data application for the purpose of demonstration, which would not affect  point estimates but might lead to a slight overestimate of the true study standard error \citep{li2020rerandomization}.  

Table~\ref{tab: data-analysis} gives the results of an intention-to-treat analysis. The point estimates of the six methods have very slight differences, the combined effect  of small-sample random variation and potential bias for the odds ratio, test-positive fraction, GLMM, and GEE estimators; their similarity further confirms the  results in \cite{utarini2021efficacy}.
Among all estimators, the covariate-adjusted estimator has the highest precision, while the test-positive fraction estimator is the least precise as reflected by its wider confidence interval, obtained by inverting tests. Comparing the covariate-adjusted and log-contrast estimators, we see that adjusting for prognostic baseline variables can lead to substantial precision gains. The GLMM and GEE estimates also adjusted for baseline variables, while their standard error estimates tend to be biased as shown in simulations.
\begin{table}[htbp]
\caption{Summary of data analysis. The point estimate and 95\% confidence interval are on the original scale, while the standard error is on the log scale, i.e, $SE(\widehat{\log(\lambda)})$.}\label{tab: data-analysis}
\centering
\begin{tabular}{rrrr}
  \hline
Estimator & Estimate & Standard Error &  95\% Confidence Interval  \\ 
  \hline
  Odds ratio & 0.23 & 0.29 & (0.13, 0.40)  \\ 
Test-positive fraction & 0.23 & - & (0.07, 0.45) \\
  GLMM & 0.24 & 0.18 & (0.17, 0.34) \\
  GEE & 0.25 & 0.17 & (0.18, 0.35) \\
   Log-contrast & 0.26 & 0.21 & (0.18, 0.40)  \\ 
  \textbf{Covariate-adjusted} & 0.25 & 0.16 & (0.18, 0.34) \\ 
   \hline
\end{tabular}
\end{table}

When considering partial compliance, we adopt the linear dose-response model and use the WEI score, defined in Section~\ref{sec: dose-response}, as the actual intervention received. The rate parameter $\beta$ is estimated at $-3.42$ (location of maximized p-value) with 95\% confidence interval $(-4.56, -2.34)$,  implying an improved intervention effect as ``dose'' increases. In the observed dose range $(0.22,0.75)$ given in Section~\ref{sec: dose-response}, $p\beta/100$ can be interpreted as the change of logarithm of relative risk per $p\%$ increase of the WEI score, assuming the linear model is correctly specified.

\section{Discussion}
\label{sec:discussion}
Since its introduction, the CR-TND has attracted increasing attention as a cost-efficient and convenient design for cluster-randomized designs to assess the effectiveness of interventions. Building on the fundamental work by \cite{anders2018cluster, jewell2019analysis}, we re-examined the current assumptions and methods for a CR-TND, and presented a new approach that allows for cluster variation in relative participant recruitment. Our proposed estimator, the log-contrast estimator, eliminates the bias that may occur in existing methods and can improve precision by adjusting for cluster-level covariates. Furthermore, we extend our results to handle partial compliance and a stepped-wedge design. 

Our proposed approaches are based on  cluster-level information, i.e., case counts, cluster-level covariates, and compliance data at the cluster level. When available, individual-level data can be summarized into cluster-level data and then analyzed by our proposed methods. Alternatively, one can explore individual-level analyses using GLMM and GEE for individual covariate adjustment (described in Section~\ref{subsec: model-based-methods}) and individual-level instrumental variable methods (\citealp{small2008randomization, clarke2012instrumental}). The validity of GLMM and GEE, however, relies on strong model assumptions; furthermore, \cite{su2021model} and \cite{wang2021robustness} both showed that an individual-level analysis can be less precise than a cluster-level analysis in cluster-randomized trials. Individual-level instrumental variable methods remains a topic for future research including extension to CR-TNDs. 

When dealing with partial compliance, we used a parsimonious linear model for the dose-response relationship since the number of clusters is small and the doses have limited coverage over the $[0,1]$ interval. For discovering a more complex dose-response relationship, a kink or spline model could be used, while they would in general need substantially more clusters or individual-level compliance data.

For the stepped-wedge design, when the intervention effect varies across clusters  by intervention start time, and by the duration of intervention, we can borrow  methodology from \cite{roth2021efficient} that established  theory for estimating a series of estimands in a traditional stepped-wedge design; in addition, their covariate adjustment techniques and finite-sample asymptotic theory can also be adapted for test-negative designs. 

{\small
\bibliographystyle{apalike}
\bibliography{references}
}


\end{document}